\documentclass{IEEEcsmag}
\pdfoutput=1
\usepackage[colorlinks,urlcolor=blue,linkcolor=blue,citecolor=blue]{hyperref}

\usepackage{upmath}
\usepackage{listings}
\usepackage{xcolor}
\usepackage{doi}

\definecolor{petsc_dark_grey}{HTML}{404243}
\definecolor{petsc_green}{HTML}{bacc33}
\definecolor{petsc_red}{HTML}{bc3b3a}
\definecolor{petsc_teal}{HTML}{6ca39e}
\definecolor{petsc_orange}{HTML}{d86035}
\definecolor{petsc_blue}{HTML}{305e89}
\definecolor{listing_bg}{HTML}{ffffff}
\definecolor{frame_color}{HTML}{CCCCCC}
\jvol{XX}
\jnum{XX}
\paper{8}
\jmonth{December}
\jname{Computing in Science and Engineering}
\pubyear{2021}

\begin{document}

%\sptitle{Department: Head}
%\editor{Editor: Name, xxxx@email}

\title{The PETSc Community Is the Infrastructure}

\author{Mark Adams}
\affil {{Lawrence} Berkeley National Laboratory, Berkeley, CA 94720, USA}

\author{Satish Balay, Oana Marin, Lois Curfman McInnes, Richard Tran Mills, Todd Munson, Hong Zhang, Junchao Zhang}
\affil{Argonne National Laboratory, Lemont, IL 60439, USA}

\author{Jed Brown}
\affil{University of Colorado at Boulder, Boulder, CO 80309, USA}

\author{Victor Eijkhout}
\affil{The University of Texas at Austin, Austin, TX 78712, USA}

\author{Jacob Faibussowitsch}
\affil{University of Illinois at Urbana-Champaign, Urbana, IL, 61801, USA}

\author{Matthew Knepley}
\affil{University of New York at Buffalo, Buffalo, NY 14260, USA}

\author{Fande Kong}
\affil{Idaho National Laboratory, Idaho Falls, ID 83415, USA}

\author{Scott Kruger}
\affil{Tech-X Corporation, Boulder, CO 80303, USA}

\author{Patrick Sanan}
\affil{ETH Zurich, 8092 Zürich, Switzerland; Argonne National Laboratory, Lemont, IL 60439, USA}

\author{Barry F. Smith}
\affil{Flatiron Institute, New York, NY 10010, USA}

\author{Hong Zhang}
\affil{Illinois Institute of Technology, Chicago, IL 60616, USA}

\markboth{}{Paper title}

\begin{abstract}
The communities who develop and support open source scientific software packages are crucial to the utility and success of such packages. Moreover, these communities form an important part of the human infrastructure that enables scientific progress. This paper discusses aspects of the PETSc (Portable Extensible Toolkit for Scientific Computation) community, its organization, and technical approaches that enable community members to help each other efficiently.
\end{abstract}

\maketitle
%
%Mark Adams,
%Satish Balay, Oana Marin, Lois Curfman McInnes, Richard Tran Mills, Todd Munson, Hong Zhang, Junchao Zhang,
%Jed Brown,
%Victor Eijkhout,
%Jacob Faibussowitsch,
%Matthew Knepley,
%Fande Kong,
%Scott Kruger,
%Patrick Sanan,
%Barry F. Smith, Hong Zhang

To meet the technological challenges of the 21st century, the simultaneous revolutions in data science and computing architectures need to be mirrored by a revolution in scientific simulation that provides flexible, scalable, multiphysics multiscale capabilities in both traditional and new areas. This simulation technology rests on a 
foundation of numerical algorithms and software for high-performance computing (HPC). This foundation rises in importance to the level of classical hard infrastructures, such as an energy grid, semiconductor foundry, or particle accelerator, but its funding and organizational models are different. Simulation technology must stand the test of time, support diverse interests, and incorporate 
cutting-edge research while running on the most advanced hardware---thus requiring significant investment for development and upgrades. Even more important, simulation technology requires human investment and new ways of organizing the effort for software and algorithm development, support, and maintenance. 

Much simulation technology today is developed and supported by using a {\it community}\footnote{Not only is the source code available but also it is developed in a public environment with contributors from a variety of institutions.} {\it open source} software paradigm
\cite{beyond, avelino2019abandonment, turk2013scale, bangerth2013makes}.
%LCM: removed 'basement' due to max refs 12
Many numerically oriented open source projects, including SciPy, Julia, and Stan, thrive because of their communities; those without a community tend to die out, have only a fringe usership, or are maintained (as orphan software) by other communities because of their importance. 

In addition, explicit focus on software ecosystems---collections of interdependent products whose development teams have incentives to collaborate to provide aggregate value---is addressing growing HPC complexity \cite{SWEcosystems:NCS2021}. Notable efforts include the xSDK, where community policies (\url{https://xsdk.info/policies}) are helping with coordination among numerical packages, and E4S (\url{https://E4S.io}), a broader effort addressing functionality across the HPC software stack. These endeavors are funded by the U.S. Department of Energy (DOE) within the Exascale Computing Project (ECP, \url{https://exascaleproject.org}). Moreover, communities are tackling challenges in {\em how} research software is developed and sustained \cite{ResearchSoftwareScienceWorkshop2021} as well as software stewardship \cite{doe-software-stewardship-RFI2021}.

This paper presents a case study of the Portable, Extensible Toolkit for Scientific Computation (PETSc) \cite{petsc-user-ref}, considering {\em community as infrastructure}. PETSc began in the early 1990s at Argonne National Laboratory as a project for research on parallel numerical algorithms. Since then, developers, users, and functionality have grown substantially, driven by continually expanding community needs to fully exploit advances in HPC architectures for next-generation science. 
PETSc consists of software infrastructure (code and tools) plus human infrastructure: the community of people who develop, support, maintain, use, and fund PETSc, their interactions, and their culture. The human infrastructure---people and their interactions as a community, within the broader DOE, HPC, and computational science communities---is foundational and enables the creation of sustainable software infrastructure. 

PETSc was not purposefully designed to support long-term community software infrastructure. Rather, work on the software inspired the creation of a set of practices to enable a small development team with large ambitions and a long time horizon to develop and support software capable of solving problems of interest to the developers and their collaborators. However, these practices, reviewed below, have wider benefits, and certain community properties could serve as a template for long-term software infrastructure:

\begin{itemize}
\item Enabling swift, in-depth engagement, especially for new users
\item Encouraging and offering opportunities for anyone to contribute to the software and documentation
\item Providing a virtual institution for collaboration
\item Developing extensible interfaces that enable people interested in mathematics and algorithms to experiment and deploy research
\item Supporting developer autonomy to pursue topics aligned with individual research needs
\item Enabling strong ties to academia, industry, and laboratories worldwide
\item Committing to continually advance library capabilities as needed by next-generation science and high-performance computer architectures 
\end{itemize}
Spread throughout the world, the PETSc community allows the real-time transfer of knowledge across institutions and application fields. Also, community interactions promote algorithmic development, enable state-of-the-art advances, and benefit the scientific community at large. 

We organize this paper as follows. In the next section, we discuss the purposes of the PETSc community and the various roles that members play. After that, we introduce several key organizational principles and communication patterns. We then introduce the paradigm of {\em debugging by email}, which encapsulates the philosophy and software technologies we use to help each other (regardless of location) use, debug, and improve PETSc.

\section{The Community}
 This section outlines the myriad purposes of the PETSc software and roles within the PETSc community. Of course, the PETSc community (and other package communities) are embedded in the DOE, HPC, and broader computational science communities. 

\paragraph{Purposes of PETSc} PETSc serves many purposes as a software library
that connects research in applied mathematics to usage within applications in science and engineering. These include the following.
\begin{itemize}
\item A \emph{research platform} targeting cutting-edge algorithmic development
\item A well-supported \emph{HPC library}
\item A \emph{repository of template applications} via a wealth of example codes
\item A \emph{compendium of algorithms}, with an algorithmic management system that provides concrete, scalable implementations of a wide range of methods described in the applied mathematics literature
\item An \emph{application development framework}
\item A \emph{pedagogical tool} for training numerical analysts on HPC platforms \cite{bueler2020petsc} 
\item A \emph{source of best-choice numerical methods} in its role as an interface between academic algorithmic development and the needs of users in science and engineering
\item An extensible \emph{interface to complementary HPC software}, such as SuperLU and hypre 
\end{itemize}

\paragraph{Roles of PETSc community members} 
Virtually all active PETSc community members are PETSc users; a smaller subset of these, who often began as PETSc users, are also PETSc developers.
All PETSc users provide important contributions, including bug reports, bug fixes, improved documentation, and suggestions for new features.
Individuals often move between different roles in the PETSc community. There is a ``long tail'' of members who contribute less frequently than the most active developers, yet collectively contribute a great deal. Over 100 people are contributors to the PETSc Git repository.

The community structure is crucial in providing a pathway to increasing involvement for interested users.
One way is to characterize PETSc community members along institutional lines.
\begin{itemize}
\item \textbf{Research-oriented users} are students, faculty, and staff focusing on research and development, who typically are employed by universities and research laboratories. 
Students may use PETSc to do homework or develop a paper or thesis code. 
Students often contribute code back to PETSc, and as they graduate, bring PETSc to their new institutions. 
Some students have become PETSc developers. 
\item \textbf{Industrial users} employ PETSc in their company's research or commercial products. 
PETSc's 2-clause BSD license eases its commercial use. These users may request support that is unlikely to be funded by research grants, 
such as support for Microsoft Windows; fortunately, there are avenues for PETSc community members to help with these requests. Industrial users require discretion and confidentiality. They cannot always share their use cases, so we must provide general solutions without details of the specifics. Developing the trust needed for industrial users is a slow, gradual process whose importance must be recognized by both sides.
\end{itemize}

\begin{figure}
 \begin{center}
  \includegraphics[trim=240 100 100 120,clip,width=0.47\textwidth]{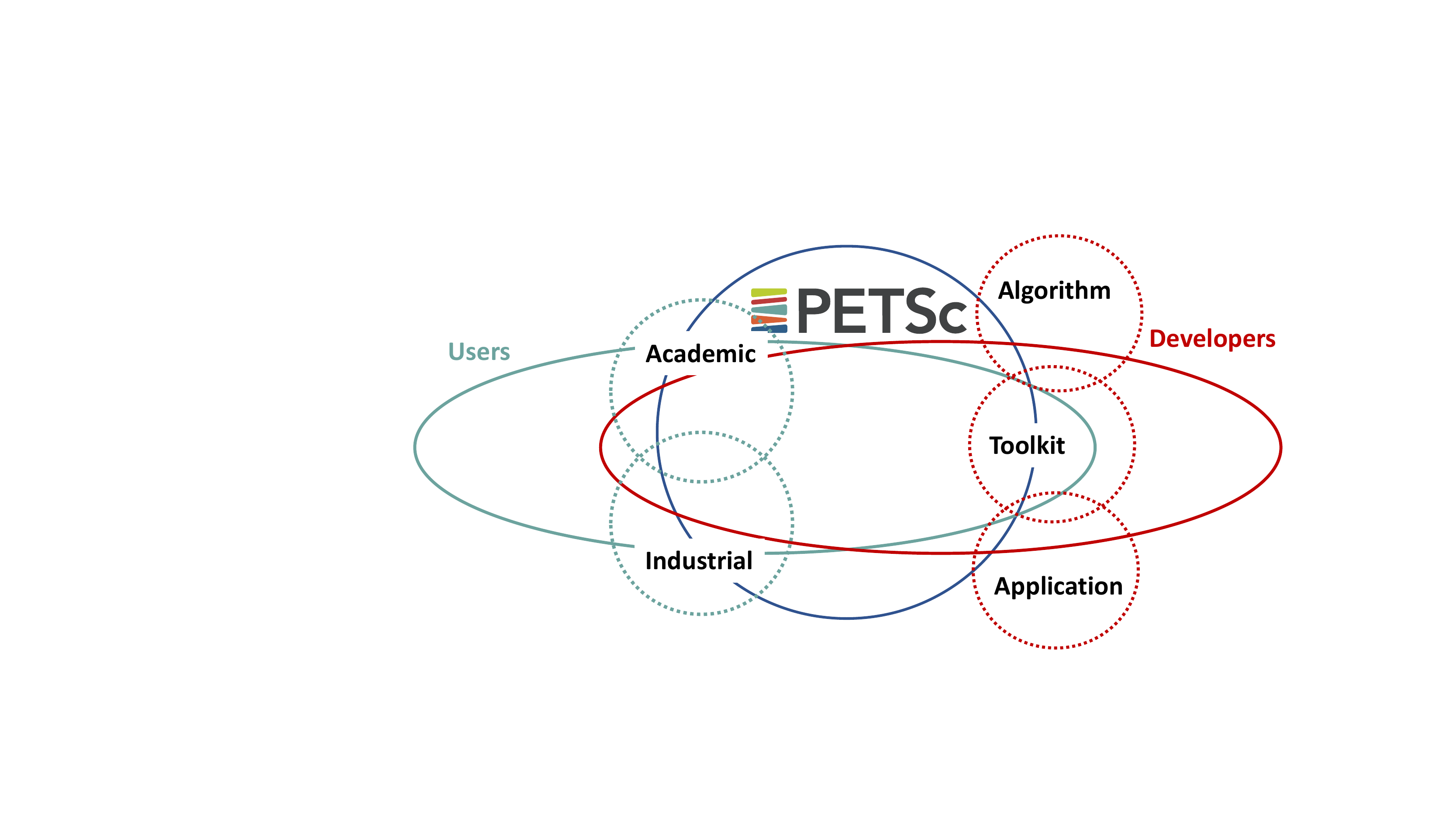}
 \end{center}
\caption{\label{fig:petsccommunity}PETSc is a scientific junction, where the boundary between types of users and developers is fluid; roles shift and change. The PETSc community has strong representation in laboratories, academia, and industry.}
\end{figure}

Another way to categorize PETSc community members is by the goals of their work, as shown in Figure~\ref{fig:petsccommunity}.
\begin{itemize}
\item \textbf{Algorithm developers} focus on devising and analyzing algorithms and hence may be less concerned about generality and usability. They use PETSc because it provides infrastructure for HPC architectures, 
allowing them to avoid unnecessary coding.
Algorithm developers face the challenge of writing scalable implementations, which, even in PETSc, can be time-consuming with a steep learning curve. However, such people find the benefits of using PETSc, including the broad impact of their work on HPC applications, outweigh using packages with a less steep learning curve, such as MATLAB.
\item \textbf{Application developers} focus on creating a code that addresses one type of simulation well, at a low cost. They are often discipline scientists or engineers, who benefit from performance enhancements provided through PETSc composability, where upgrades in algorithms and data structures can occur seamlessly from a user perspective, yet provide significant performance increases. 
\item \textbf{Scientific toolkit developers} build systems that tackle a subset of PETSc functionality but at a higher level of abstraction, with more specific support for their target class of problems.
Such toolkits, including Firedrake, MOOSE, and Deal.II, leverage PETSc capabilities and introduce additional infrastructure.
\end{itemize}

\section{Organization and Communication}

We now summarize the organizational and communication patterns that have emerged in PETSc due to its various purposes and member roles.
Besides communicating within funded projects, institutional settings, and events, users engage in the PETSc through online support mailings, GitLab issues, and Slack channels. Since 2015, annual PETSc user meetings have included tutorials on how to leverage library functionality for research while also highlighting users' science achievements made possible by advances in PETSc features.

\paragraph{Engagement when problems occur}
Support is a crucial aspect of a healthy software community.
Members of the PETSc community usually respond to requests within hours, if not minutes.
This engagement helps new users feel welcomed and valued, make rapid progress, and gain confidence. 
Through users' feedback, experienced community members learn what works, what doesn't, and where improvements are needed.
PETSc community members often discover new research topics from feature requests and discussions.

However, providing excellent support is a substantial effort, particularly when users encounter difficult bugs or performance issues at scale.
User-developer communication on a particular topic can span weeks or even months.
PETSc developers need to be patient and consistently engaged with users. One may question whether this practice is sustainable, but it has worked reasonably well. In the last section, we discuss some technical approaches to providing support.

\paragraph{Trust within the community}
To maintain the vitality of the library, new algorithmic developments must be rapidly integrated, bugs promptly fixed, and awkward constructions removed. These activities require the PETSc community to establish a high level of trust, communicating that the library will be well supported even in the face of rapid evolution, and that code will continue to run with help from the community. Establishing trust is a precondition to creating a welcoming environment for new users and developers. Members of the PETSc community have a wide range of professions, backgrounds, and levels of involvement, with individuals often participating in several ways over the years. Engagement is key to disseminating tacit knowledge and developing users' skills and social support, so that people can transition to become developers and mentors. The PETSc community has developed a broad base of people with expertise and kindness 
to reduce and report bugs, mentor newcomers, and contribute in other ways. 

\paragraph{Community to community}
Application communities often treat PETSc as a software ecosystem instead of a stand-alone package. As a result, they rely on PETSc to manage necessary low-level tools, such as MPI, BLAS/LAPACK, and vendor packages used on accelerators. Application communities also appreciate unified solver interfaces, particularly linear preconditioners, which enable application codes to access third-party libraries such as MUMPS, SuperLU, and hypre with little effort. 
Often ``technical language'' barriers exist between communities.  For example, a contact mechanics expert may describe a solver convergence issue as ``we found a PETSc error when contact mechanics occurs with a frictionless model." A numerical solvers expert might find it hard to resolve such an issue. Fortunately, other PETSc community members may have domain expertise and can serve as liaisons between communities.
Such individuals speak the languages of both communities, understanding both PETSc's capabilities and the needs of their communities. Thus, they can appropriately explore, explain, and introduce PETSc features to their communities.
Such liaisons help expand PETSc's reach across disciplines and reduce the centralized maintenance burden by addressing many questions directly in their communities while contributing patches, feature requirements, and even serving as testers of software releases.

An important PETSc subcommunity is system engineers who manage institutions' computational infrastructure. They often are the first to encounter problems that need the attention of PETSc developers. Their expertise can help rapidly debug problems and develop fixes that need to be quickly merged into the distribution. Package maintainers, for example, for {\em APT}, are also a valuable resource, as they track PETSc on particular configurations; they often have excellent suggestions for improvements to PETSc's configuration and installation. 

\paragraph{Responding to change} Communities must be able to respond with innovative and creative solutions to changing circumstances. For numerical software, this includes the continual emergence of new science drivers and techniques, including data science and artificial intelligence, as well as new hardware architectures. For example, a large shift in HPC is currently underway with incorporating graphical processing units (GPUs) into both high-end and moderate-scale scientific computing. Major organizations such as DOE have responded with, for example, the Exascale Computing Project, where community open source projects, including PETSc, are aggressively developing innovations in data structures and algorithms for new architectures \cite{MILLS2021}, \cite{PetscSF2022}. The PETSc community empowers developers to be creative by providing the autonomy to be innovative while still maintaining guidelines for development~\cite{PETScdevelopersguide} and town squares to organize overall development plans. This approach, along with PETSc's wide variety of contributors, enables a level of agility that might not otherwise occur. This approach also naturally promotes project-specific planning (for example, as needed for work proposed and funded in particular grants) and coordination among development communities overall. 

\paragraph{Enabling research collaborations} PETSc's community helps members identify funding opportunities, access expertise, and transition between roles in the project. One of the greatest difficulties in maintaining a coherent software project over decades is providing career paths for contributors. PETSc gives academic contributors a solid foundation for advancement via awards (e.g., SC Gordon Bell prizes, SIAM prizes), the highly cited users’ manual, professional recognition, and productive collaborations born from PETSc development, maintenance, and support. In addition, PETSc provides resource sharing from collaborative grants and collaboration opportunities that extend beyond the development group. Sometimes, PETSc affiliation may be more important than departmental affiliation, especially since modern academic departments are often atomized, with little internal collaboration. The community provides strong academic connections for industrial and laboratory members, tangible outputs recognized by future employers, and active participation in the wider computational science community.

\paragraph{Engagement with funding institutions} Even the smallest community open source projects cannot exist without some funding and institutional support. Usually, it is a combination of grants from governmental or non-governmental agencies, in-house funding within particular institutions, and less formal systems that allow employees to contribute to open source packages during a portion of their regular employment. Members of the PETSc community are actively engaged with program development, including communicating with program managers at the U.S. Department of Energy and the National Science Foundation and with management at their institutions, to ensure that support is provided and maintained. This form of interaction is crucial to the long-term viability of all open source software communities; PETSc users and community members have played important roles in various local, national, and international conversations, including recent DOE activities related to software sustainability
\cite{ResearchSoftwareScienceWorkshop2021,doe-software-stewardship-RFI2021}. 

\section{Debugging by Email} \label{sec:debug}

Helping people when they encounter problems is essential for the PETSc community.
Members of the community need to quickly diagnose where problems
occur and refer users to corresponding support if a solution is beyond
the expertise of PETSc developers.
 Because of complexity---users call PETSc through its Python, 
Fortran, or C bindings; in Linux, macOS, or Windows operating systems; on machines from laptops
 to the world's most powerful supercomputers; on x86, Arm, Power, or other CPU architectures, possibly accelerated by 
GPUs from different vendors---PETSc community helpers cannot reproduce all problems met by individuals.
Thus, PETSc has a holistic remote debuggable design, enabling a feature dubbed \textit{debugging by email},
 which means that PETSc community members can pinpoint 
 causes of problems through conversations in PETSc mailing lists and GitLab issues.
 
\begin{figure}
 \begin{center}
 \includegraphics[trim=240 90 90 60,clip,width=0.6\textwidth]{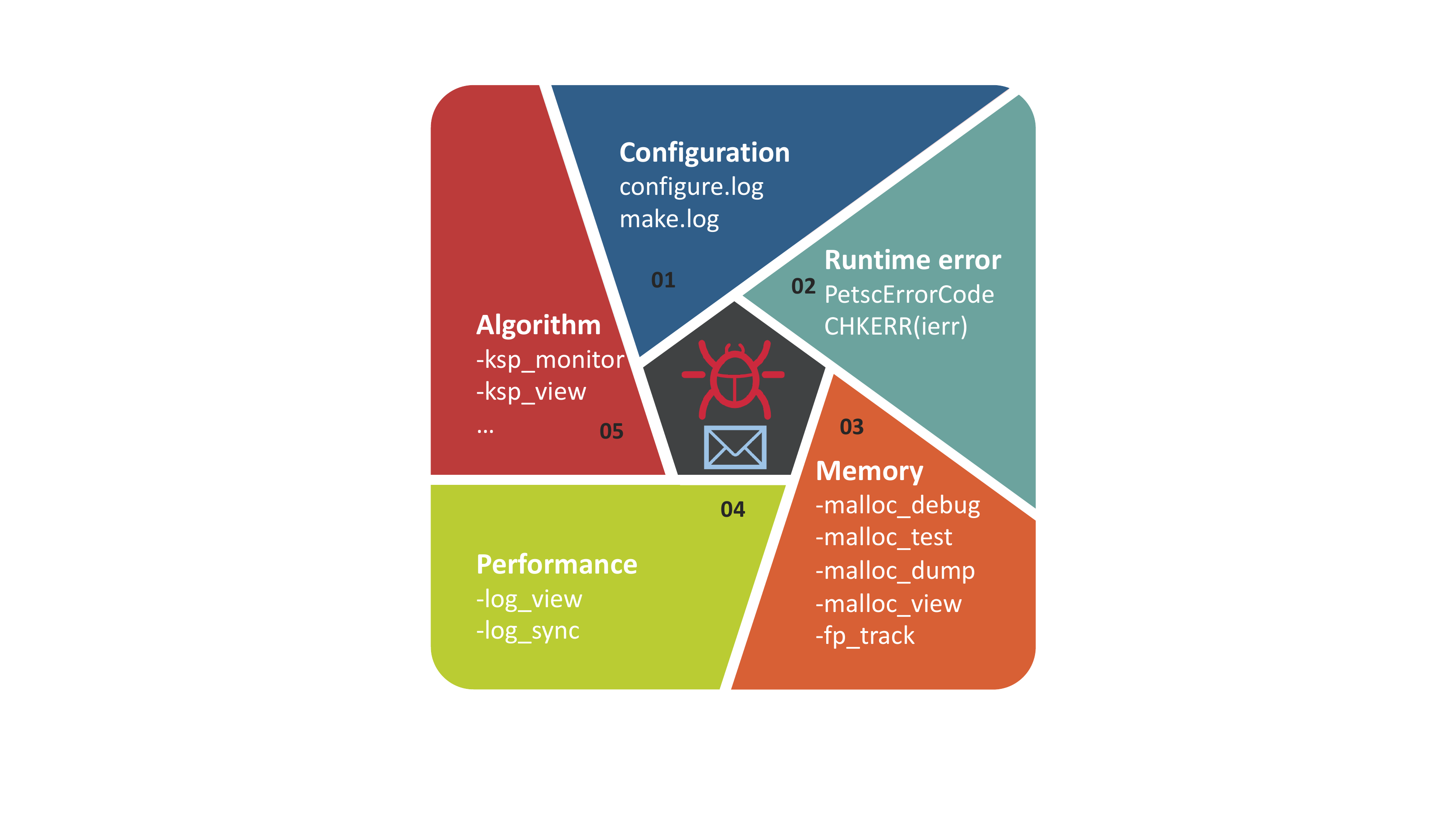}
 \end{center}
\caption{\label{fig:debug}PETSc has a holistic remote debuggable design, enabling a feature dubbed \textit{debugging by email},
 which means that PETSc community members can pinpoint 
 causes of problems through conversations in PETSc mailing lists and GitLab issues.}
\end{figure}

This approach 
enables the community to do more with less.
In this section, we introduce the debuggability
design of PETSc, as it closely affects engagement and support.
PETSc's debuggability spans from configuration to code execution, with the salient features highlighted in Figure \ref{fig:debug}. 
Our commitment to software support leads us to maintain our own simple, 
integrated tools for many tasks that conventional software wisdom would 
commonly dictate should be performed by full-featured external tools.

\paragraph{Configuration debugging}
HPC software systems have complex execution environments,
including great variance in hardware and software.
Software developers must expend considerable effort 
to configure and build their code for different situations. Then, if failures occur,
the software developers have to diagnose them.
Configuration failures are the most common support issues the PETSc community faces.
Thus, having a debuggable configuration system with comprehensive logging is critical.
Rather than using a standard configuration system such as GNU Autotools or CMake,
PETSc has a bespoke configuration system with extensive checking, written in Python, 
 which logs everything during 
configuration in a single file \texttt{configure.log}.
When a check fails, it generates
clear error messages and a Python stack trace.
PETSc users attach \texttt{configure.log} and another file \texttt{make.log} generated
by \texttt{make} when they meet configuration errors.
By examining the two files, PETSc developers can quickly determine why the configuration system 
made specific choices and what went wrong.
In addition, since the configuration system is bespoke, the PETSc community can easily add new checking, testing, and logging. 
CMake is notoriously difficult to debug by email because it logs information in various directories and does not log much of its process.

\begin{lstlisting}[
  keywordstyle=\color{petsc_green},
  stringstyle=\color{petsc_blue}\ttfamily,
  stringstyle=\color{petsc_teal}\ttfamily,
  identifierstyle=\color{petsc_blue},
  commentstyle=\color{petsc_green},
  emphstyle=\color{petsc_red},
  language=C,
  basicstyle=\ttfamily\footnotesize, 
  numbers=left,
  float=*,
  label={lst:vecaxpy}, 
  caption={Sample showing error checking in PETSc\vspace{5pt}}
]
PetscErrorCode  VecAXPY(Vec y,PetscScalar alpha,Vec x){
  PetscErrorCode ierr;
  PetscFunctionBegin;
  PetscValidHeaderSpecific(x,VEC_CLASSID,3);
  PetscValidType(x,3);
  ...
  PetscCheckSameTypeAndComm(x,3,y,1);
  VecCheckSameSize(x,3,y,1);
  if (x == y) SETERRQ(comm,PETSC_ERR_ARG_IDN,"x and y cannot be the same vector");
  PetscValidLogicalCollectiveScalar(y,alpha,2);
  ...
  ierr = PetscLogEventBegin(VEC_AXPY,x,y,0,0);CHKERRQ(ierr);
  ierr = (*y->ops->axpy)(y,alpha,x);CHKERRQ(ierr);
  ierr = PetscLogEventEnd(VEC_AXPY,x,y,0,0);CHKERRQ(ierr);
  ...
  PetscFunctionReturn(0);
\end{lstlisting}

\paragraph{Runtime error debugging}
The PETSc library strives to provide descriptive error messages that explain why and where errors have occurred, making it easy for PETSc developers
to diagnose by email what went wrong and assist users with fixes.
PETSc has extensive code to assist in this regard.
See Listing \ref{lst:vecaxpy},
which shows code that adds two vectors with \texttt{y += alpha*x}.
Every PETSc function returns a \texttt{PetscErrorCode}, indicating
whether the function is successfully executed, and if not, 
what type of error occurred. 
In PETSc source code, every function call is error-checked, in a style similar to lines 12--14.
We make errors manifest early rather than later to avoid obscure error messages. 
The default error handler
prints the stack trace leading to the error, 
including function names, file names, and line numbers.
The stack trace is built inside the two macros \texttt{PetscFunctionBegin}
and \texttt{PetscFunctionReturn(0)}; see lines 3 and 16. 
PETSc also provides utilities to check the integrity of function parameters; see lines 4--10.
All application programming interfaces (APIs) shown here are public; users are encouraged to apply the same strategy in their code.

PETSc also has APIs to assert properties of the code so that 
useful error messages are generated promptly if the code behaves unexpectedly.
For example, once a matrix is preallocated or assembled, one can 
set a property of the matrix to
 indicate that in subsequent insertions 
one will insert only to existing nonzero locations.

\paragraph{Memory debugging}
Memory corruption problems are common; therefore, memory allocations are done through a PETSc-specific API that 
records information and sets sentinels around the allocations in debug mode.
With command-line options, PETSc will initialize
the allocated memory with NaN (not a number); 
using the uninitialized memory
in floating-point operations will generate an appropriate error message.
Also, PETSc can check the integrity of the entire heap of PETSc-allocated
memory at every allocation.
PETSc codes also can output information about memory that has never been freed during the PETSc finalization stage,
to detect memory leaks.
As lightweight Valgrind-like features, the output can be shared with PETSc developers to help understand a code's misbehavior. But, of course, we also recommend using more sophisticated tools, including Valgrind and debuggers.

\paragraph{Performance debugging}
Another challenging support task, which the PETSc community also handles routinely, 
is debugging performance problems, particularly for high levels of parallelism.
PETSc provides APIs
to set stages in their code and log the performance of events of interest.
For example, lines 12 and 14 in Listing \ref{lst:vecaxpy}
are for the \texttt{VEC\_AXPY} event, rendering
a lightweight, integrated logging system that allows users to quickly 
gather timings.
Listing \ref{lst:logview} shows a snippet of the stdout output. 
From the top, we know that, besides the PETSc's default stage (Main Stage), the test created
 another stage (Solve).
PETSc summarizes the computation and communication statistics of the two stages.
Below that, it lists detailed statistics of events (usually functions) within each stage (only stage 1 is shown here),
including the number of times an event has been called, time and floating-point operations (flops) an event has spent,
MPI messages and reductions an event has incurred, and other statistics. 
Interested readers are referred to the PETSc/TAO Users Manual \cite{petsc-user-ref} 
or hints in the log view message itself.
Because MPI processes typically have different statistics in parallel,
PETSc shows maxima over all processes and ratios of maxima to minima. 
This information is useful because whenever we encounter a large ratio in time or flops in output, we know
a load imbalance in the corresponding event might exist.
Sometimes, imbalance in one event can distort the timing of other events 
(for instance, processes might wait for
messages from a lagging partner), giving confusing results.
\begin{lstlisting}[
  float=*,
  basicstyle=\ttfamily\tiny, 
  label={lst:logview},
  caption={Sample output of -log\_view\vspace{5pt}}
]
Summary of Stages:   ----- Time ------  ----- Flop ------  --- Messages ---  -- Message Lengths --  -- Reductions --
                        Avg     %Total     Avg     %Total    Count   %Total     Avg         %Total    Count   %Total
 0:      Main Stage: 2.3986e-03  33.6%  4.6080e+03   0.6%  1.200e+01   1.9%  9.300e+02        8.4%  2.300e+01   9.1%
 1:           Solve: 4.7290e-03  66.3%  8.1830e+05  99.4%  6.200e+02  98.1%  1.963e+02       91.6%  2.100e+02  83.3%

Phase summary info:
   Count: number of times phase was executed
   Time and Flop: Max - maximum over all processors
                  Ratio - ratio of maximum to minimum over all processors
   Mess: number of messages sent
   AvgLen: average message length (bytes)
   Reduct: number of global reductions
   Global: entire computation
   Stage: stages of a computation. Set stages with PetscLogStagePush() and PetscLogStagePop().
      %T - percent time in this phase         %F - percent flop in this phase
      %M - percent messages in this phase     %L - percent message lengths in this phase
      %R - percent reductions in this phase
   Total Mflop/s: 10e-6 * (sum of flop over all processors)/(max time over all processors)
------------------------------------------------------------------------------------------------------------------------
Event                Count      Time (sec)     Flop                              --- Global ---  --- Stage ----  Total
                   Max Ratio  Max     Ratio   Max  Ratio  Mess   AvgLen  Reduct  %T %F %M %L %R  %T %F %M %L %R Mflop/s
------------------------------------------------------------------------------------------------------------------------
MatMult               54 1.0 4.4847e-04 1.5 8.86e+04 1.1 3.4e+02 1.6e+02 1.0e+00  5 42 53 39  0   8 42 54 43  0   762
MatMultAdd             5 1.0 7.3937e-05 4.3 2.13e+03 1.1 1.5e+01 9.1e+01 0.0e+00  1  1  2  1  0   1  1  2  1  0   111
MatMultTranspose       5 1.0 4.6906e-05 2.2 2.23e+03 1.1 2.1e+01 7.2e+01 1.0e+00  0  1  3  1  0   1  1  3  1  0   179
VecAXPY               48 1.0 9.4440e-06 1.8 9.60e+03 1.0 0.0e+00 0.0e+00 0.0e+00  0  5  0  0  0   0  5  0  0  0  4066
VecScatterBegin       73 1.0 2.6231e-04 1.6 0.00e+00 0.0 4.5e+02 1.6e+02 5.0e+00  3  0 72 53  2   4  0 73 58  2     0
VecScatterEnd         73 1.0 4.4319e-04 3.6 1.70e+02 0.0 0.0e+00 0.0e+00 0.0e+00  4  0  0  0  0   6  0  0  0  0     0
KSPSetUp               5 1.0 6.2018e-04 2.8 4.53e+04 1.0 6.0e+01 1.6e+02 3.6e+01  5 22  9  7 14   7 22 10  8 17   288
KSPSolve               1 1.0 1.0645e-03 1.0 1.26e+05 1.0 2.4e+02 1.5e+02 1.4e+01 15 60 38 27  6  22 61 39 29  7   466
\end{lstlisting}

 Having the profiling tools integrated with the numerical algorithms in use,
 outputting by default to \texttt{stdout}, is crucial because it allows all users to provide quickly 
 information on their usage, independent of what computational
 systems they may use or which additional analysis or logging tools they have available.
 This allows PETSc developers to quickly and directly view timings on the user's system and facilitate performance debugging of scalable solvers 
 at ``production'' scale, by email, where direct reproduction of a user's issue is infeasible. 

\paragraph{Algorithm debugging}
PETSc includes an extensive suite of parallel preconditioners, 
linear solvers, nonlinear solvers,
and time integrators.
Composable and nested solvers are among the most powerful PETSc features since they facilitate numerical experimentation on novel, highly complex problems, but keeping track of them can be difficult.
PETSc developers must see the detailed solver configurations to spot potential problems. Hence, PETSc provides APIs to display all solver options being used.
Listing \ref{lst:snesview} shows a snippet of a longer output from a nonlinear solver.
With indention reflecting levels of the composite solvers, we can clearly 
see the nested solvers used and key parameters employed at various levels of the solvers.
%snes_tutorials-ex56_0
\begin{lstlisting}[
  float=*,
  basicstyle=\ttfamily\tiny, 
  label={lst:snesview},
  caption={Sample output of -snes\_view\vspace{5pt}},
]
SNES Object: 4 MPI processes
  type: newtonls
  maximum iterations=2, maximum function evaluations=10000
  tolerances: relative=1e-10, absolute=1e-50, solution=1e-08
  total number of linear solver iterations=14
  total number of function evaluations=2
  norm schedule ALWAYS
  SNESLineSearch Object: 4 MPI processes
    type: bt
      interpolation: cubic
      alpha=1.000000e-04
    maxstep=1.000000e+08, minlambda=1.000000e-12
    tolerances: relative=1.000000e-08, absolute=1.000000e-15, lambda=1.000000e-08
    maximum iterations=40
  KSP Object: 4 MPI processes
    type: cg
    maximum iterations=100, initial guess is zero
    tolerances:  relative=1e-10, absolute=1e-50, divergence=10000.
    left preconditioning
    using UNPRECONDITIONED norm type for convergence test
  PC Object: 4 MPI processes
    type: gamg
      type is MULTIPLICATIVE, levels=3 cycles=v
        Cycles per PCApply=1
        Using externally compute Galerkin coarse grid matrices
        GAMG specific options
          Threshold for dropping small values in graph on each level = 0.05  0.  0.    
          ...
          Complexity:    grid = 1.05401
    Coarse grid solver -- level -------------------------------
      ...
      PC Object: (mg_coarse_) 4 MPI processes
        type: bjacobi
          number of blocks = 4
          ...
\end{lstlisting}
PETSc provides flexible monitors to be used in conjunction with solver views, 
that print the residual or function norm at each iteration of an iterative solver so that users
can check the convergence of the solver and compare different algorithms. 
Listing \ref{lst:snesmonitor} shows output of nonlinear and linear solver monitors.
\begin{lstlisting}[
  float=*,
  basicstyle=\ttfamily\tiny, 
  label={lst:snesmonitor},
  caption={Sample output of -snes\_monitor -ksp\_monitor\vspace{5pt}}
]
  0 SNES Function norm 1.223958326481e+02 
    0 KSP Residual norm 1.223958326481e+02 
    1 KSP Residual norm 2.137523917735e+01 
    2 KSP Residual norm 4.364326343132e+00 
    ...
   12 KSP Residual norm 3.922463112223e-10 
  Linear solve converged due to CONVERGED_RTOL iterations 12
  1 SNES Function norm 3.922318262147e-10 
Nonlinear solve converged due to CONVERGED_FNORM_RELATIVE iterations 1
\end{lstlisting}

In summary, PETSc offers a wide set of complementary options to aid {\em debugging by email} with the following common themes:
users can enable debugging
regardless of their computing environment;
errors appear as early as possible; and
output is printed in well-formatted plain text for copy-and-paste or file attachments
to emails and GitLab issues.

\section{Conclusion}
The increased prominence of data science and the transition to computing architecture heterogeneity require more, not less, high-quality numerical simulation and analysis software. This software is often created in community open source environments; the communities are crucial to the utility of such software. We have briefly outlined some aspects of the open-source PETSc community and its collaboration strategies. However, most of what was discussed applies broadly to other numerical software communities. We concluded by focusing on mechanisms we use to allow community members to efficiently help one other at a distance using straightforward communication channels. The science and engineering of scientific software communities is only just beginning, and this topic is starting to receive more consideration at institutional levels. By sharing some of the PETSc community approaches, we hope to contribute to the wider scientific computing community as it seeks to improve the software programming process. 

\section*{Acknowledgments}
The authors thank all PETSc users and developers for their many software, organizational, and conceptual contributions to the community. 
This material was partially based upon work funded by the U.S. Department of Energy, Office of Science, under contract DE-AC02-06CH11357. 
This work was supported by the Exascale Computing Project (17-SC-20-SC), a collaborative 
effort of the U.S. Department of Energy Office of Science and the National Nuclear Security Administration, and by the U.S. Department of Energy, Office of Science, Office of Advanced Scientific Computing Research under Award Number DE-SC0016140. 
Matt Knepley and Jed Brown were partially supported by U.S. DOE Contract DE-AC02-0000011838.

\bibliographystyle{plain}
\bibliography{mybibfile}

\begin{IEEEbiography}{Mark Adams}{\,}
Mark Adams is a staff scientist in the Scalable Solvers Group at Lawrence Berkeley National Laboratory. Contact him at mfadams@lbl.gov.
\end{IEEEbiography}

\begin{IEEEbiography}{Satish Balay}{\,}is a software engineer at Argonne National Laboratory (ANL). Contact him at balay@mcs.anl.gov.
\end{IEEEbiography}

\begin{IEEEbiography}{Oana Marin}{\,}is with ANL, as a numerical analyst. Contact her at oanam@anl.gov.
\end{IEEEbiography}

\begin{IEEEbiography}{Lois Curfman McInnes}{\,}is an Argonne Distinguished Fellow. Contact her at curfman@anl.gov.
\end{IEEEbiography}

\begin{IEEEbiography}{Richard Tran Mills}{\,} is a computational scientist at ANL. Contact him at rtmills@anl.gov.
\end{IEEEbiography}

\begin{IEEEbiography}{Todd Munson}{\,}is a senior computational scientist at ANL. Contact him at tmunsonr@anl.gov.
\end{IEEEbiography}

\begin{IEEEbiography}{Hong Zhang}{\,}is an assistant computational mathematician at ANL. Contact him at hongZhang@anl.gov.
\end{IEEEbiography}

\begin{IEEEbiography}{Junchao Zhang}{\,}is a software developer at ANL. Contact him at jczhang@anl.gov.
\end{IEEEbiography}

\begin{IEEEbiography}{Jed Brown}{\,}is an assistant professor at the University of Colorado at Boulder. Contact him at jed@jedbrown.org.
\end{IEEEbiography}

\begin{IEEEbiography}{Victor Eijkhout}{\,}is a research scientist at the Texas Advanced Computing Center of the University of Texas at Austin. Contact him at eijkhout@tacc.utexas.edu.
\end{IEEEbiography}

\begin{IEEEbiography}{Jacob Faibussowitsch}{\,}is a graduate research student at the University of Illinois at Urbana-Champaign. 
%He is a Ph.D. student in the department of Mechanical Sciences and Engineering.
Contact him at faibuss2@illinois.edu.
\end{IEEEbiography}

\begin{IEEEbiography}{Matthew Knepley}{\,}is an associate professor at the University of New York at Buffalo. Contact him at knepley@gmail.com.
\end{IEEEbiography}

\begin{IEEEbiography}{Fande Kong}{\,}is a computational scientist and software developer at Idaho National Laboratory.
%He received his Ph.D. in computer science from the University of Colorado Boulder in 2016. Currently, he is a MOOSE developer.
Contact him at fande.kong@inl.gov.
\end{IEEEbiography}

\begin{IEEEbiography}{Scott Kruger}{\,}is a scientist/VP with Tech-X Corporation. Contact him at kruger@txcorp.com.
\end{IEEEbiography}

\begin{IEEEbiography}{Patrick Sanan}{\,}is a postdoctoral researcher at ETH Zurich and assistant computational mathematician at ANL. 
%He received his PhD in Applied and Computational Mathematics from the California Institute of Technology. He is interested in scalable solvers and scientific software engineering. 
Contact him at psanan@anl.gov.
\end{IEEEbiography}

\begin{IEEEbiography}{Barry F. Smith}{\,}is a senior research scientist at the Flatiron Institute; he was previously a Senior Computational Mathematician at ANL. 
%He received a Ph.D. in Applied Mathematics at the Courant Institute in 1990. 
Contact him at bsmith@flatironinstitute.org.
\end{IEEEbiography}

\begin{IEEEbiography}{Hong Zhang}{\,}is a research professor of computer science at the Illinois Institute of Technology. Contact her at hzhang@mcs.anl.gov
\end{IEEEbiography}

%TC:ignore

\onecolumn
\centering
\framebox{
\parbox{4in}{
The submitted manuscript has been created by UChicago Argonne, LLC, Operator of Argonne
National Laboratory (``Argonne''). Argonne, a U.S. Department of Energy Office of Science
laboratory, is operated under Contract No. DE-AC02-06CH11357. The U.S. Government retains
for itself, and others acting on its behalf, a paid-up nonexclusive, irrevocable worldwide
license in said article to reproduce, prepare derivative works, distribute copies to the
public, and perform publicly and display publicly, by or on behalf of the Government.
The Department of Energy will provide public access to these results of federally
sponsored research in accordance with the DOE Public Access
Plan. \url{http://energy.gov/downloads/doe-public-accessplan}
}}

%TC:endignore
\end{document}